\documentclass[epsfig,graphics,twocolumn,floatfix,prl]{revtex4}\usepackage{graphicx}

\begin{document}\title[a]{A solid state light-matter interface at the single photon level}

\author{Hugues de Riedmatten \footnotemark[1]\footnotetext{\footnotemark[1] These authors
contributed equally to this work.}}\author{Mikael Afzelius
\footnotemark[1]}\author{Matthias U. Staudt}\author{Christoph
Simon}\author{Nicolas Gisin}

\address{Group of Applied Physics, University of Geneva, CH-1211 Geneva 4, Switzerland}

\begin{abstract}
Coherent and reversible mapping of quantum information between
light and matter is an important experimental challenge in quantum
information science. In particular, it is a decisive milestone for
the implementation of quantum networks and quantum repeaters
\cite{Briegel1998,Duan2001,Simon2007}. So far, quantum interfaces
between light and atoms have been demonstrated with atomic gases
\cite{Julsgaard2004,Chaneliere2005,Eisaman2005,Chou2005,Honda2008,Appel2008},
and with single trapped atoms in cavities \cite{Boozer2007}. Here
we demonstrate the coherent and reversible mapping of a light
field with less than one photon per pulse onto an ensemble of
$\sim10^{7}$ atoms naturally trapped in a solid. This is achieved
by coherently absorbing the light field in a suitably prepared
solid state atomic medium \cite{Afzelius2008}. The state of the
light is mapped onto  collective atomic excitations on an optical
transition and stored for a pre-programmed time up of to 1$\mu s$
before being released in a well defined spatio-temporal mode as a
result of a collective interference. The coherence of the process
is verified by performing an interference experiment with two
stored weak pulses with a variable phase relation. Visibilities of
more than 95$\%$ are obtained, which demonstrates the high
coherence of the mapping process at the single photon level. In
addition, we show experimentally that our interface allows one to
store and retrieve light fields in multiple temporal modes. Our
results represent the first observation of collective enhancement
at the single photon level in a solid and open the way to
multimode solid state quantum memories as a promising alternative
to atomic gases.

\end{abstract}
 \maketitle

Efficient and reversible mapping of quantum states between light
and matter requires strong interactions between photons and atoms.
With single quantum systems, this regime can be reached with high
finesse optical cavities, which is technically highly demanding
\cite{Boozer2007}. In contrast, light can be efficiently absorbed
in ensembles of atoms in free space. Moreover, it is possible to
engineer the atomic systems such that the stored light can be
retrieved in a well defined spatio-temporal mode due to a
collective constructive interference between all the emitters.
This collective enhancement is at the heart of protocols for
storing photonic quantum states in atomic ensembles, such as
schemes based on Electromagnetically-Induced Transparency
(EIT)\cite{Fleischhauer2000}, off-resonant Raman interactions
 \cite{Duan2001,Nunn2007}  and modified photon echoes using
Controlled Reversible Inhomogeneous Broadening
(CRIB)\cite{Moiseev2001,Nilsson2005,Kraus2006} and Atomic
Frequency Combs (AFC) \cite{Afzelius2008}.

All previous quantum storage experiments with ensembles have been
performed using atomic gases as the storage material
\cite{Julsgaard2004,Chaneliere2005,Eisaman2005,Chou2005,Honda2008,Appel2008}.
However, some solid state systems have properties that make them
very attractive for applications in quantum storage. In
particular, rare-earth ion doped solids provide a unique physical
system where large ensembles of atoms are naturally trapped in a
solid state matrix, which prevents decoherence due to the motion
of the atoms and allows the use of trapping free protocols.
Moreover, these systems also exhibit excellent coherence
properties at low temperature (below 4 K), both for the optical
\cite{Bottger2006a} and spin transition \cite{Fraval2005}. The
long optical coherence times enable storage of multiple temporal
modes in a single QM, which promises significant speed-up in
quantum repeater applications\cite{Simon2007}. Furthermore, high
optical densities can be obtained in rare-earth doped solids,
which is required to achieve strong light matter coupling
resulting in high efficiency light storage and retrieval. However,
despite recent experimental progress
\cite{Longdell2005,Alexander2006,Hetet2008,Staudt2007a,Staudt2007,Ohlsson2003a},
the implementation of a solid state light matter quantum interface
has not been reported so far. \\


\begin{figure*}    \centering    \includegraphics[width=.80\textwidth]{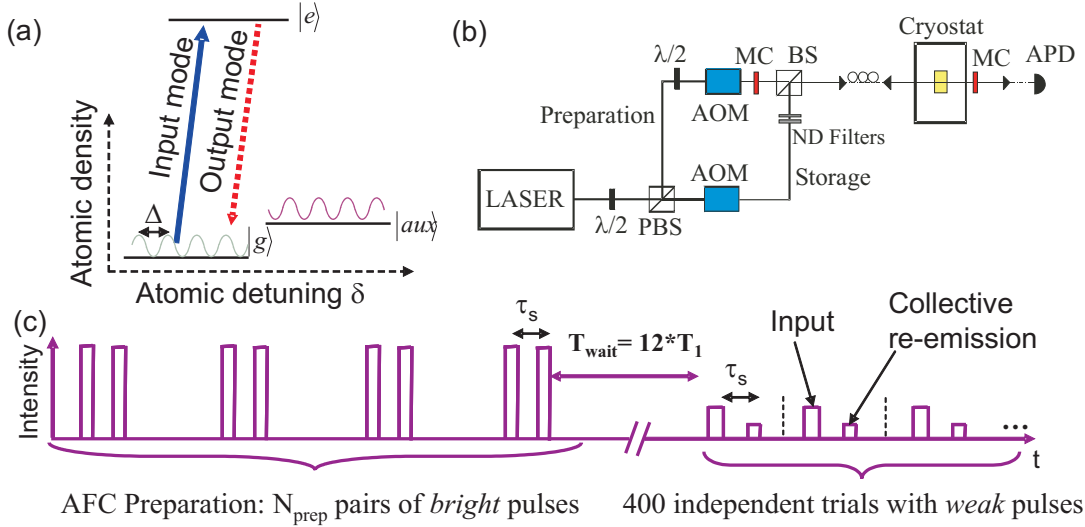}
\caption{\footnotesize(a) Solid state light matter interface. The
light is absorbed on the $^4I_{9/2}\rightarrow ^4F_{3/2}$
transition of Nd$^{3+}$ ions
 at 880 nm. The inhomogeneous broadening of the optical transition is 2 GHz, with a maximal optical depth around 4.
 The optical relaxation time $T_1$ of the excited state
 $|e\rangle$ is equal to 100 $\mu s$. The ground state is split into two Zeeman levels $|g\rangle$ and $|aux\rangle$,
 separated by 3.8 GHz through the application of a magnetic field of 300
 mT. A spectral grating is prepared in $|g\rangle$ by a preparation
 sequence described in the Methods section.
 (b) Experimental setup. The laser
source is a cw external cavity diode laser at 880 nm. The laser is
split at a variable beam splitter and the pulse sequences for the
preparation of the grating and for the pulses to be stored are
prepared by independent acousto-optics modulators in different
optical paths. The duration of the pulses is about 30 ns. These
paths are then recombined at a beam splitter (BS) and coupled into
a single mode fiber to ensure proper mode matching. The light is
then focussed onto 1 mm long Nd$^{3+}$:YVO$_4$ crystal with a beam
diameter of 30 $\mu m$. The crystal is cooled down to 3 K by a
pulse tube cooler. After the sample, the light passes through a
polarizer and is coupled back to a single mode fiber, which is
connected to a Silicon Avalanche Photo diode single photon
counter. In order to block the preparation light during the
storage and to protect the detector from the intense preparation
light, two mechanical choppers (MC) are used. (c) Optical pulse
sequence. The experimental sequence is divided into two parts: the
preparation of the spectral grating (see Methods) and the storage
of the weak pulses. We wait a time $T_w =12T_1=1200 \mu s$ between
the preparation and the storage sequence in order to avoid
fluorescence. During the storage sequence, 400 independent trials
are performed at a repetition rate of 200 kHz. The entire sequence
preparation plus storage is then repeated with a repetition rate
of 40 Hz.} \label{setup}\end{figure*}

Here we demonstrate for the first time the coherent and reversible
mapping of weak coherent states of light with less than one photon
per pulse onto a large number of atoms in a solid. The mapping is
done by coherently absorbing the light in an ensemble of
inhomogeneously broadened atoms spectrally prepared with a
periodic modulation of the absorption profile
\cite{Mossberg1979,Carlson1984,Mitsunaga1991,Afzelius2008}. The
reversible absorption by such a spectral grating is at the heart
of the recently proposed multimode quantum memory scheme based on
AFC \cite{Afzelius2008}. We therefore also demonstrate a proof of
principle of an essential primitive of this protocol at the single
photon level.

 Let us now describe in more detail how our interface works.
Assume an incident weak coherent state of light $|\alpha\rangle_L$
with a mean photon number $\overline{n}=|\alpha|^2<1$. After
absorption the photons are stored in a coherent superposition of
collective optical excitations de-localized over all the atoms in
resonance with the light field. The state of the atoms (not
normalized) can be written
as:\begin{equation}|\alpha\rangle_{A}=|0\rangle_A+\alpha|1\rangle_A+O(\alpha^2)\end{equation}where
$|0\rangle_A=|g_1\cdot\cdot\cdot g_N \rangle$ and

\begin{equation}|1\rangle_A=\sum_i c_i e^{i\delta_i
t}e^{-ikz_i} |g\cdot\cdot\cdot e_i\cdot\cdot\cdot g \rangle
\end{equation}

where $z_i$ is the position of atom $i$ (for simplicity, we only
consider a single spatial mode defined by the direction of
propagation of the input field), $k$ is the wave-number of the
light field, $\delta_i$ the detuning of the atom with respect to
the laser frequency and the amplitudes $c_i$ depend on the
frequency and on the spatial position of the particular atom $i$.
This collective state will rapidly dephase since each term
acquires an individual phase $e^{i\delta_i t}$ depending on the
detuning. However, due to the periodic structure of the absorption
profile, the collective state will be re-established after a
pre-programmed time $2\pi/\Delta$, where $\Delta$ is the period of
the spectral grating. This leads to a coherent photon-echo type
re-emission in the forward spatial mode
 \cite{Mossberg1979,Carlson1984,Mitsunaga1991,Afzelius2008}. Note
that in our experiment the light field is stored as a collective
excitation on the optical transition, contrary to all previous
experiments at the single photon level, where collective
excitations of spin states were used
 \cite{Chaneliere2005,Eisaman2005,Chou2005}. Conceptually, this is
an important difference since in our case the light is simply
absorbed in the prepared material, without any control field.

The solid state interface is implemented in an ensemble of
Neodymium  ions (Nd$^{3+}$) doped into a YVO$_4$ crystal
\cite{Hastings-Simon2008}. The Nd$^{3+}$ ions constitute an
ensemble of inhomogeneously broadened atoms having a relevant
level structure with two spin ground states $|g\rangle$ and $|aux
\rangle$ and one excited state $|e\rangle$, as shown in
Fig.\ref{setup}. Initially, the two ground states are equally
populated for all frequencies over the inhomogeneous broadening.
The preparation of the spectral grating is realized by frequency
selective optical pumping from $|g\rangle$ to $|aux\rangle$ via
the excited state $|e\rangle$ (see Fig.\ref{setup} for an overview
of the experiment). This is implemented with a Ramsey type
interference using a train of coherent pairs of pulses
\cite{Hesselink1979}
 (see Fig. \ref{setup} and Methods). To store weak light
fields, it is required that there is no population in the excited
state, otherwise fluorescence will blur the signal. This is
ensured by waiting long enough between the preparation and storage
sequences, such that all atoms have returned to the ground states.
As a result of the preparation sequence, a spectral grating is
present in $|g\rangle$ before the storage begins, which decays
with the population relaxation lifetime $T_Z$ between the spin
states ($T_Z$ = 6 ms in our case).

\begin{figure}    \centering    \includegraphics[width=.50\textwidth]{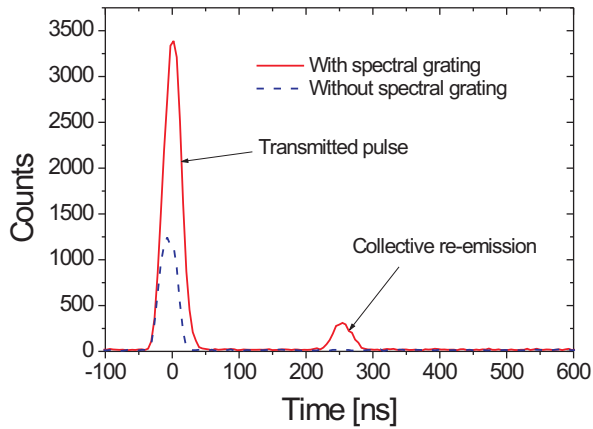}

\caption{\footnotesize Reversible mapping of a coherent state with
$\overline{n}=0.5$.  The solid line corresponds to the case where
a spectral grating is prepared with a periodicity of 4 MHz. The
peak 250 ns after the transmitted input pulse corresponds to the
collective retrieval after storage in the solid state medium. The
dashed line corresponds to the case where the atomic medium is not
prepared with a spectral grating. In that case, we only see the
transmitted input pulse  (about 2 percent of the incoming pulse is
transmitted). The absorption of the input pulse is smaller when
the spectral grating is prepared (about 5 percent of the light is
transmitted)}. \label{echo}\end{figure}

\begin{figure}\centering\includegraphics[width=.50\textwidth]{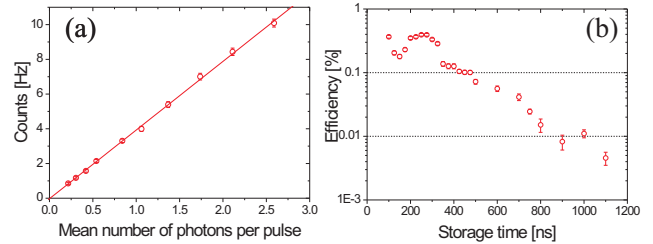}
\caption{\footnotesize (a) Number of detections in the collective
output mode as a function of $\overline{n}$. (b) Efficiency as a
function of the storage time for $\overline{n}$ = 2.7. For small
storage times ($<$400 ns), an oscillation of the efficiency is
clearly visible. This a quantum beat due to the interaction of the
electronic spin of the Nd$^{3+}$ ion with the nuclear spin of the
surrounding Vanadium ions (called super hyperfine interaction).
For longer storage times, the decay is exponential with decay
constant of 220 ns. This interaction also limits the minimal width
of the absorption peaks.} \label{number}\end{figure}

In the first experiment, we demonstrate collective mapping of weak
coherent states $|\alpha\rangle_L$ on the crystal. An example with
$\overline{n}$ =0.5 is shown in Fig. \ref{echo} (See Methods for
the estimation of $\overline{n}$). When the sample is prepared
with a spectral grating having a periodicity of 4 MHz, we observe
a strong emission at the expected storage time of 250 ns. About
0.5 percent of the incoming light is re-emitted in this signal.
This is more than 4 orders of magnitude more than what would be
expected from a non collective re-emission, taking into account
that we collect a solid angle of $2\cdot10^{-4}$ and that the
optical relaxation time is 3 orders of magnitude longer than the
observed signal. This signal thus clearly arises from a collective
re-emission, which demonstrates the collective and reversible
mapping of a light field with less than one photon onto an large
number of atoms in a solid. To further study the mapping process,
we record the number of counts in the observed signal for
different $\overline{n}$ ranging from 0.2 to 2.7 (see Fig.
\ref{number} a). This shows that the mapping is linear and that
very low photon numbers can still be mapped and retrieved. We also
investigated the decay of storage efficiency with the storage time
(see Fig. \ref{number}b).


The efficiency of the storage and retrieval and its decay as a
function of storage time can be qualitatively understood using the
theory of the AFC quantum memory \cite{Afzelius2008}. In order to
obtain high storage efficiencies for a given storage time, it is
essential to create a spectral grating with narrow absorption
peaks as compared to the spectral separation of the peaks, i.e. a
high finesse grating \cite{Afzelius2008}.  In our experiment,
however, the minimal width of the absorption peak is of the order
of 1-2 MHz, due to material properties (see Fig.\ref{number}) and
to the linewidth of our free-running laser. Hence the grating is
close to a (low finesse) sinusoidal curve, which limits the
storage efficiency and causes the observed decay of the efficiency
vs storage time. Moreover, there is still a significant flat
absorption background due to imperfect optical pumping in the
present experiment. This background can be considered as a loss
which strongly limits the observed storage efficiency. We
emphasize however that these are not fundamental limitations of
rare-earth materials. The preparation of narrow absorption lines
with low absorption background has been demonstrated for several
other rare-earth materials, such as Eu:Y$_2$SiO$_5$
\cite{Alexander2006}, Pr:Y$_2$SiO$_5$ \cite{Hetet2008,Rippe2008}
and Tm:YAG \cite{Seze2003}. Significantly higher storage
efficiencies should be obtained in these materials
\cite{Afzelius2008}. Note also that the experiment described here
implements a memory device with a fixed storage time. In order to
allow for on-demand read-out of the stored field (as required for
quantum repeaters) and storage times longer than that given by the
spectral grating, the
 excitations in $|e\rangle$ can be transferred to a
third ground state spin level $|s\rangle$. This transfer also
enables in principle storage efficiencies close to unity
\cite{Afzelius2008}.

\begin{figure}    \centering
\includegraphics[width=.50\textwidth]{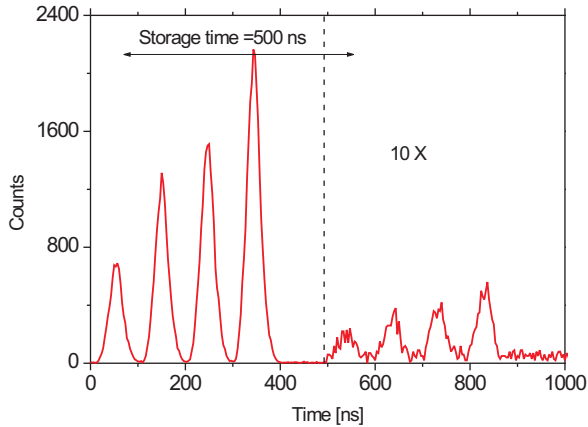}
\caption{\footnotesize Multimode light matter interface. The
spectral grating is prepared for a storage time of 500 ns. The
first four pulses are the transmitted pulses. After 500 ns, we can
clearly see the collective re-emission of the four temporal modes.
 The signal to noise ratio is smaller than for the single
mode case (fig. \ref{echo}), due to the longer storage time of 500
ns (cf Fig.\ref{number} b). For clarity, the output signal part
has been magnified by a factor of ten.}
\label{fig_prot}\end{figure}

So far we have considered the storage of weak light fields in a
single temporal mode. However the use of spectral gratings also
allows for the storage in multiple temporal modes as shown in
\cite{Afzelius2008}. The maximal number of modes that can be
stored is given by the ratio of the storage time (determined by
the spectral grating) to the duration of an individual mode. To
illustrate this multimode property we store trains of four weak
pulses with $\overline{n}$ from 0.8 to 0.3 during 500 ns, as shown
in Fig.\ref{fig_prot}. It is important to note that the time
ordering of the pulses is preserved during the storage, which
results in the same storage efficiency for each mode. This is in
contrast with CRIB based quantum memories where the time ordering
is reversed \cite{Kraus2006}. The number of stored modes for a
given storage time can be improved by using shorter preparation
pulses, resulting in larger bandwidth of the spectral grating. In
this case, shorter pulses can be stored. In the present
experiment, the shortest duration of pulses was set to about 20 ns
(FWHM) by technical limitations. A great advantage of the AFC
protocol is that the number of modes that one can store does not
depend on the optical depth, contrary to EIT and CRIB
\cite{Afzelius2008}.

\begin{figure}
\centering
\includegraphics[width=.50\textwidth]{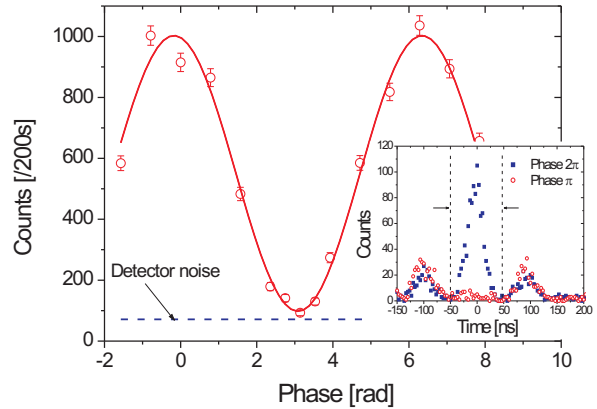}
\caption{\footnotesize Interference fringe. Time-bin qubits with
different phases $\Phi$ are stored and analyzed using the
interface. The analysis is performed by projecting the time-bin
qubit on a fixed superposition basis, which here is achieved by
two partial read-outs (see text for details). The inset shows the
histogram of arrival times, where there is a constructive ($\Phi$
= 2$\pi$) and destructive interference ($\Phi$ = $\pi$) in the
middle time bin. For this particular interference fringe, we
obtain a raw visibility of 82$\%$, or 95$\%$ when subtracting
detector dark counts.} \label{fringe}\end{figure}

For applications in quantum memories, it is crucial that the
interface conserves the phase of the incoming pulses. To probe the
coherence of the mapping process at the single photon level, we
use a pair of weak pulses separated by a time $\tau$ = 100 ns with
a fixed relative phase $\Phi$. This can be seen as a time bin
qubit, which can be written as:
 $|\psi\rangle_L=|\alpha_t\rangle_L+e^{i\Phi}|\alpha_{t+\tau}\rangle_L$
where $|\alpha_t\rangle_L$ represents a weak coherent state at
time t. The qubit is stored and thereafter analyzed directly in
the memory using a method developed in Ref \cite{Staudt2007a}.
This method requires the implementation of partial read-outs at
different times. This realizes a projection on a superposition
basis, similarly to what can be done with a Mach-Zehnder
interferometer \cite{Staudt2007a}. If the time $\tau$ between the
weak pulses matches the time between the two read-outs, the
re-emission from the sample can be suppressed or enhanced
depending on the phase difference between the two incoming pulses.
The visibility of this interference is a measure of the coherence
of the mapping process. The two partial read-outs are here
achieved by preparing two super-imposed spectral gratings having
different periods, corresponding to storage times of 200 and 300
ns. An example of interference for $\overline{n}$ =0.85 is shown
in Fig.\ref{fringe}. Dark-counts subtracted visibilities above 95
$\%$ have been obtained for various $\overline{n}$ between 0.4 and
1.7, which demonstrates the high coherence of the storage process,
even at the single photon level. From these visibilities, one can
infer conditional fidelities above 97 $\%$ for the storage of
single photons \cite{Staudt2007}. This excellent phase
preservation at the single photon level is obtained thanks to the
collective enhancement effect \cite{Staudt2007} and to the almost
complete suppression of background noise .

In conclusion, we have demonstrated the coherent and reversible
mapping of a light field at the single photon level onto a solid.
Our results show that the storage of single photons (Fock states)
in multiple temporal modes in solids is possible. We have also
demonstrated that the quantum coherence of the incident weak light
fields is almost perfectly conserved during the storage. Solid
state systems can therefore be considered as a promising
alternative to atomic gases for photonic quantum storage. This
line of research holds promise for the implementation of efficient
long distance
quantum networks. \\

\textbf{Methods}

\footnotesize The preparation of the spectral grating is realized
by a series of pairs of pulses of area $\Theta<\pi/2$ resonant
with the $|g\rangle \rightarrow |e\rangle$ transition. Each pair
of pulses realizes a frequency selective coherent transfer of
population from $|g\rangle$ to $|e\rangle$ \cite{Hesselink1979}.
This can be seen as a Ramsey type interference where the two light
pulses play the role of beam splitters and the phase shift
acquired in the excited state depends on the detuning of the
atoms. The periodicity of the created spectral grating is then
given by the inverse of the time interval $\tau_s$ between the two
pulses. The atoms in the excited state can decay to both ground
states $|g\rangle$ and $|aux\rangle$ with a relaxation time $T_1$=
100 $\mu s$. The atoms that decay to $|aux\rangle$ are not
affected by the preparation laser and remain in this state for a
time of 6 ms \cite{Hastings-Simon2008}. The pulse sequence is then
repeated 100 times with a time separation between the pairs of 15
$\mu s$, longer than the optical coherence time $T_2$=7 $\mu s$.
This allows for the build up of the spectral grating, with
population storage in $|aux\rangle$.

To estimate the mean number of photons per pulse, we shift the
laser out of resonance with the absorbing atoms and record the
proportion of detections in the single photon counter (typically
between $1 \%$ and $20 \%$). By a careful measurement of the
detection efficiency ($\eta_D$=0.32) and of the transmission
efficiency from the input face of the cryostat to the detector
(typically $\eta_t$=0.2), one can finally infer the mean number of
photons in front of the cryostat, before the sample.

\textbf{Acknowledgements} We thank E. Cavalli and M.Bettinelli for
kindly lending us the Nd:YVO$_4$ crystal. This work was supported
by the Swiss NCCR Quantum Photonics and by the European Commission
under the Integrated Project Qubit Applications (QAP).


\end{document}